\documentclass{aa}
\usepackage{graphicx}
\begin{document}
   \title{Serendipitous discovery of seven new southern L-dwarfs}
   \author{T.R. Kendall
          \inst{1}
          \and
          N. Mauron\inst{2}
          \and 
          M. Azzopardi\inst{3}
          \and
          K. Gigoyan\inst{4}\thanks{Based on observations made with the ESO Danish 1.54m telescope, 
    La Silla, program 69.B-0186(A)} }

   \offprints{T.R. Kendall}

   \institute{Centro de Astronomia e Astrof\'{\i}sica da Universidade de Lisboa,
              Observat\'{o}rio Astron\'{o}mico de Lisboa, Tapada da Ajuda,  
              1349-018 Lisboa, Portugal\\ \email{tkendall@oal.ul.pt} \and
              Groupe d'Astrophysique, UMR-5024 CNRS, Case 072, Place 
              Bataillon, F-34095 Montpellier Cedex 05, France \and
              IAM, Observatoire de Marseille, 2 Place Le Verrier, 13248 
              Marseille Cedex 04, France \and 378433 Byukaran 
              Astrophysical Observatory \& Isaac Newton Institute of 
              Chile, Armenian Branch, Ashtarak d-ct, Armenia}

    \date{received <date> /  Accepted <date>}

   \abstract{We report the discovery of seven hitherto unknown L-dwarfs
found as a result of a spectroscopic search for distant AGB stars. Their far-red and near-infrared colours
are very similar to known dwarfs of the same spectral type. One new object is among the $\sim$30 brightest L-dwarfs, with  $K_s$=12.12, and is nearby, $\sim$\,20pc. Using low resolution spectroscopy from the Danish 1.54\,m ESO telescope, 
spectral types in the range L0.5 -- L5 are derived for these seven L-dwarfs by direct comparison to L-type standards taken from \cite{kir99}. Distances are determined from
existing calibrations, and together with measured proper motions, yield kinematics for the seven
new dwarfs consistent with that expected for the solar neighbourhood disk population. 
   }
   \titlerunning{Seven new L-dwarfs}
   \authorrunning{T.R. Kendall et al.}
   \maketitle

\section{Introduction}

Recent near-infrared surveys, most notably 2MASS (\cite{skr97}) and DENIS (\cite{epc97}) are an invaluable resource for finding and characterising
the reddest existing objects. Among galactic objects, one can mention the
cool AGB stars, which may be deeply embedded in circumstellar dust,
and the L/T dwarf population. The latter such objects, when observed in the field, are likely to have ages
of a few Gyr and, with reference to theoretical models (\cite{bur97};~\cite{bar98}), to be substellar. As such,
ever since the discovery of the first such objects
in the solar neighbourhood, GD\,165B (\cite{bec88}) and Gliese 229\,B (\cite{nak95}), they have been the subject of intense observational scrutiny. Current and recent studies include \cite{del97}; \cite{del99}; \cite{kir99}; \cite{mar99}; \cite{rei00}; \cite{fan00} and \cite{sch02} and references therein.

\begin{table*}
\begin{center}
\caption[] {Basic and derived data for observed targets. The abbreviated name 2MJhhmm will be used throughout this paper. Previously known objects are referenced in the final column. Coordinates are from 2MASS 2nd incremental release. $I$ magnitudes are taken from
UKST $I$ photometry from SuperCosmos. $JHK_s$ magnitudes are from the 2MASS
database. Galactic latitudes, derived absolute $J$-magnitudes and distances (where applicable, see text) are given in cols. 7, 8 and 9. Spectral types gathered
from previous literature are given in parentheses.}  
\begin{tabular}{lllllllllll}\hline\hline
 Name & 2MASS & $I$ & $J$ & $H$ & $K_s$ & $b$ & M$_{\rm J}$ & Sp. & d/pc & ref. \\
\hline
2MJ0006 & 2MASS{\bf I} J0006205-172051 & 18.86 & 15.71 & 14.63 & 14.02 & -75.7 & 12.45 & L2.5 & 45 & - \\
2MJ0013 & 2MASS{\bf I} J0013578-223520 & 19.42 & 15.75 & 14.57 & 14.01 & -80.4 & 12.89 & L4 & 37 & - \\
2MJ0032 & 2MASS{\bf I} J0032431-223727 & 18.56 & 15.38 & 14.52 & 13.97 & -83.8 & 12.01 & L1 & 48 & - \\
2MJ0034 & 2MASS{\bf I} J0034568-070601 & 18.74 & 15.55 & 14.56 & 13.95 & -69.6 & 12.60 & L3 & 39 & - \\
2MJ0125 & 2MASS{\bf I} J0125369-343505 & 18.30 & 15.48 & 14.47 & 13.86 & -79.5 & 12.31 & L2 & 43 & - \\
2MJ0144 & 2MASS{\bf I} J0144353-071614 & 18.08 & 14.19 & 13.00 & 12.28 & -66.5 & 13.18 & (L5) & 18 & L02 \\
2MJ0205 & 2MASS{\bf I} J0205293-115930 & 17.83 & 14.58 & 13.59 & 12.98 & -70.0 & 13.76 & (L7) & 11 & D97 \\
2MJ0423 & 2MASS{\bf I} J0423486-041403 & 18.40 & 14.45 & 13.44 & 12.94 & -34.5 &  -   & (T0) & -  & G02, S02 \\
2MJ0428 & 2MASS{\bf I} J0428510-225323 & 16.96 & 13.58 & 12.70 & 12.12 & -40.9 & 11.88 & L0.5 & 22 & - \\
2MJ0443 & 2MASS{\bf I} J0443058-320209 & 17.81 & 15.27 & 14.34 & 13.87 & -40.0 & 13.18 & L5 & 26 & - \\
\hline
\end{tabular}
\end{center}
\vspace*{2mm}
\begin{flushleft}
L02: \cite{lie03}\\
D97: \cite{del97}\\
G02, S02: \cite{geb02,sch02}
\end{flushleft}
\end{table*}

It is interesting to note that during a survey devoted to the search for L and T-dwarfs, four dusty AGB carbon stars were discovered at high galactic latitude ($|b|$\,$\geq$\,25$^{\circ}$),
as reported by \cite{lie00}. Such luminous AGB stars are invaluable probes of the distant halo (see for example \cite{tot98}). In this paper, we report the inverse case: the discovery of seven new L-dwarfs found during a systematic spectroscopic survey of AGB candidates.

Our program to search for high latitude AGB stars is still in progress
and will be described fully elsewhere (\cite{mau03}). Briefly, AGB candidates are
selected either from objective prism plates (\cite{gig01}) or by using a combination of near-infrared and optical colours. The majority of them have $R$\,$\sim$\,13\,--\,19.  However, a subset of 
ten candidates which are invisible on POSS R-plates, but relatively bright 
in the 2MASS $K_s$-band ($K_s$\,=\,11--14) were immediately selected for further examination 
using deeper $R$ and $I$-band imaging, to attempt to detect the object in the $R$-band. 
Subsequently, low resolution (30\,\AA) grism spectroscopy was carried out (see below). 

Of the
ten objects followed up in this way, seven are shown to be hitherto
unknown L-dwarfs. Two were found to be in the Kirkpatrick database (version 6 December 2002)\footnote{http://spider.ipac.caltech.edu/staff/davy/ARCHIVE}, hereafter
K02, which now contains 250 objects classified as spectral type L. These are DENIS-P
J0205.4-1159 (Delfosse et al. 1997) and 2MASS J0144353-071614 (Liebert et al. 2003). A further one 
was found in recent literature by searching the object coordinates
in SIMBAD: this is SDSS J042348.57-041403.5, listed by \cite{sch02};~\cite{geb02}, classified T0.

   \begin{figure}
  \centering
  \resizebox{\hsize}{!}{\includegraphics[angle=-90]{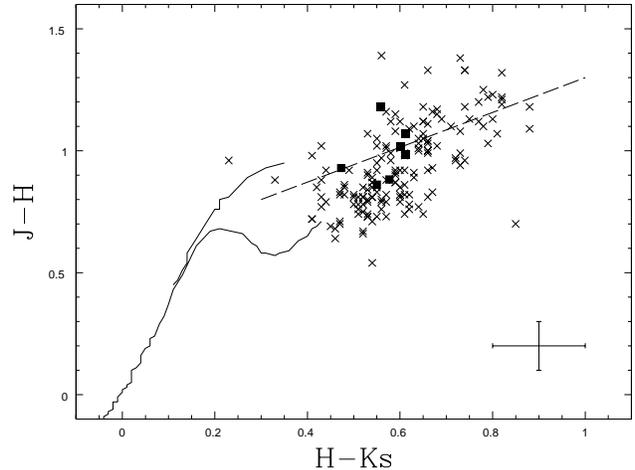}}
  \caption {Near-infrared 2MASS colour-colour diagram for the seven new L-dwarfs
(large filled squares) compared to a subset 161 known L-dwarfs from K02 (crosses) with well-defined $JHK_s$ magnitudes (errors on $J-H$ and $H-K_s$ $\leq$ 0.1 mag., the size shown by the errorbar at lower right). Errors
on the 2MASS colours for the new L-dwarfs are similar; $\sim\pm$ 0.08 mag. Also shown are the loci for main sequence (the lower line, near $H-K$\,=\,0.3) and giant stars, together with an indication of the locus where the majority of cool halo carbon stars lie (near to the dashed line).   }
  \label{jhk}
   \end{figure}

  \begin{figure}
  \centering
  \resizebox{\hsize}{!}{\includegraphics[angle=-90]{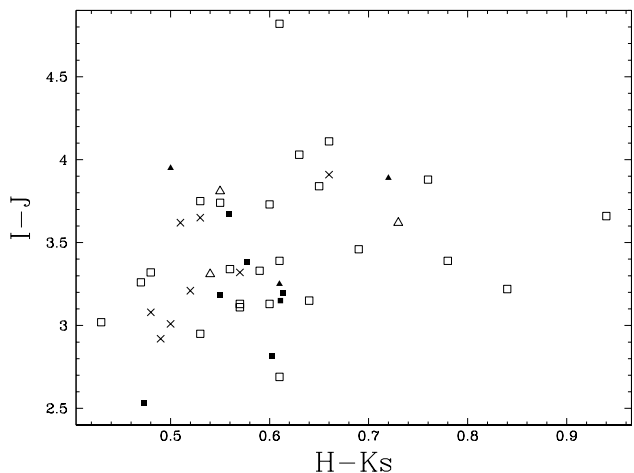}}
  \caption{$I-J$, $H-K_s$ colour-colour diagram. The 7 filled squares are the seven
new L-dwarfs, and the filled triangles are the other three known objects listed in Table 1. Open squares are 24 L-dwarfs from Kirkpatrick et al. (1999), excluding DENIS-P
J1228-1547. This object is included in a subsample of objects from Delfosse et al. (1999) which are known L-dwarfs included in K02 (open triangles). Crosses
are data from Mart\'{\i}n et al. (1999). For consistency, $JHK_s$ magnitudes have been taken from 2MASS in all cases. 
For our data, $I$ magnitudes have been taken from SuperCosmos,
while those of Mart\'{\i}n et al. (1999) and \cite{kir99} are derived from their flux calibrated spectra.
Other $I$ magnitudes are from DENIS.}
  \label{ijhk}
   \end{figure}

\section{Observations and Data Reduction}
 
The observations were carried out with the 1.54\,m Danish telescope
 at the European Southern Observatory in La Silla (Chile), from August 29
 to Sept. 3, 2002. The weather was good and the seeing varied from
 0.9 to 2\arcsec.

\begin{figure}
  \centering
  \resizebox{\hsize}{!}{\includegraphics[]{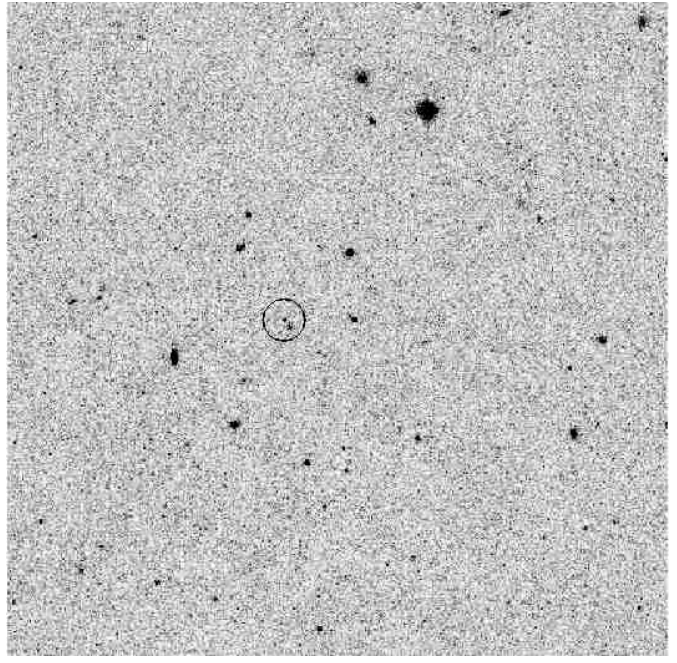}}
  \caption{$I$-band finder chart for 2MJ0125. The L-dwarf is the
object to the NE of a faint galaxy also within the circle.}
  \label{e0902}
   \end{figure}

\begin{figure}
  \centering
  \resizebox{\hsize}{!}{\includegraphics[]{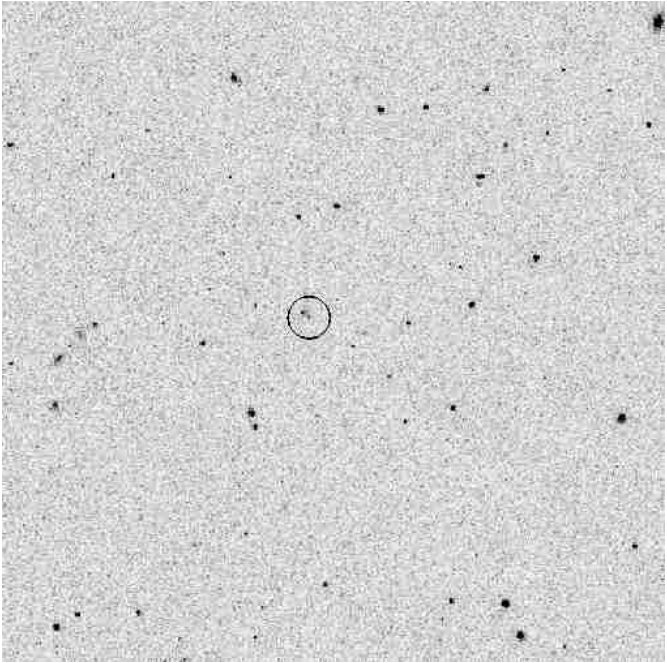}}
  \caption{$I$-band finder chart for 2MJ0443. The L-dwarf is the object to the SW of a faint galaxy, also in the circle.}
  \label{d0809}
   \end{figure}

 We used the DFOSC focal reducer which is designed to permit
 both direct imaging onto a 2048$\times$2048 CCD through a variety of filters
 and slit spectroscopy with grisms. For each program source, 2\,min exposure 
 images in Bessell $R$ and Gunn $i$ bands were first achieved 
 for source identification.
 This step is necessary to see whether the source is
 point-like or extended (and hence discarded) and to see rapidly if the counterpart is also red in $R-I$.
 This imaging is also useful to note if there is a possible confusion with
 another close object, which is the case for 2 new objects where a faint galaxy was found to be 
within a few arcsec of the target (see Figs.~\ref{e0902} and ~\ref{d0809}).
Pre-imaging also permits a rough estimate
 of the source counts in the $R$-band filter to decide whether low or medium resolution
 spectroscopy should be attempted.

 Spectroscopy was then performed with the source positioned on
 a 1.5\arcsec~ slit and DFOSC
 grisms 8 or 12, which provide dispersions of  1\,\AA\,pix$^{-1}$ and 12\,\AA\,pix$^{-1}$
 respectively, i.e. 2.5 pixel resolutions of 2.5 and 30\AA. Exposure times ranged between 10\,min and 1\,hr.
 The intermediate resolution spectra cover 5850 -- 8400\,\AA, while
the observed range at low resolution is 5600 -- 10000\,\AA.

 Imaging flat-fields in $R$ and $I$-bands
 were achieved on the telescope dome, as well as spectroscopic dome flats
 with the slit and the grisms present in the optical beam.

 Spectra of the photometric standard
 white dwarf GD50 were secured with the same 1.5\arcsec~slit as for the targets and were
 used to flux calibrate the final target spectra.
 These fluxes are of uncertain absolute accuracy, possibly
 not better than a factor of 2, although for a few objects
 comparison of two spectra taken on different conditions and on
 two different dates suggests agreement to within $\sim$\,20\%. Nevertheless, when
comparing fluxes from wavelength to wavelength, the quality of the spectra
is quite sufficient to allow comparison with calibrated templates and derive
spectrophotometric indices (see below).

 Data reduction was performed with the MIDAS software installed at
 Montpellier. After bias subtraction and flat-field correction of
 the 2-D spectra, a one-dimensional spectrum of the source was 
 extracted and the sky spectrum measured close to the source location 
 was subtracted. Cosmic ray hits were removed manually. No correction for telluric absorption 
has been performed.  

 For  wavelength calibration of the low resolution
 spectra, it was found necessary to select and use  the few strong and
 isolated (unblended at low resolution) lines provided
 by the sky spectrum, together with several of the lines provided by 
 separate He-Ne lamp exposures. Only sky lines were used for the
 medium-resolution spectrum of 2MJ0144. The residuals in the dispersion
 relation were found to be $<$1/3 pixel ($\sim$\,1/6 
 of a resolution element).
 
 The main defect of the low resolution spectra is the presence
 of interference fringes for wavelengths beyond $\sim$\,8500\,\AA.
 These fringes do not cancel out easily by flat fielding: they
 are not a spatially stable pattern in CCD spectral frames
 and vary from exposure to
 exposure due to mechanical flexure of the instrument. Therefore, 
 the spectra which are displayed here could not 
 be completely corrected. However, our conclusions do not rely heavily
on fringe-affected regions. 

\begin{figure}
  \centering
  \resizebox{\hsize}{!}{\includegraphics[]{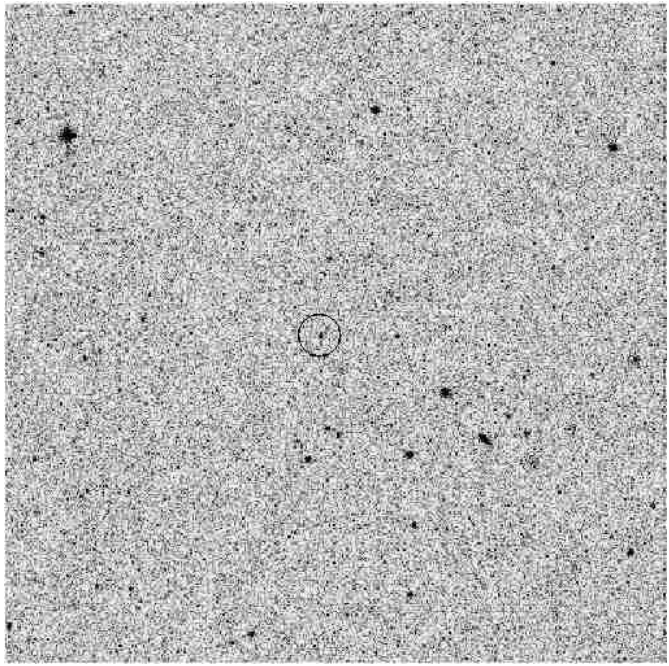}}
  \caption{$I$-band finder chart for 2MJ0006. All charts presented here are Gunn $i$-band images from the Danish 1.54\,m telescope and are 5\arcmin\, on a side: N is up, E to the left.}
  \label{e0900}
   \end{figure}

  \begin{figure}
  \centering
  \resizebox{\hsize}{!}{\includegraphics[]{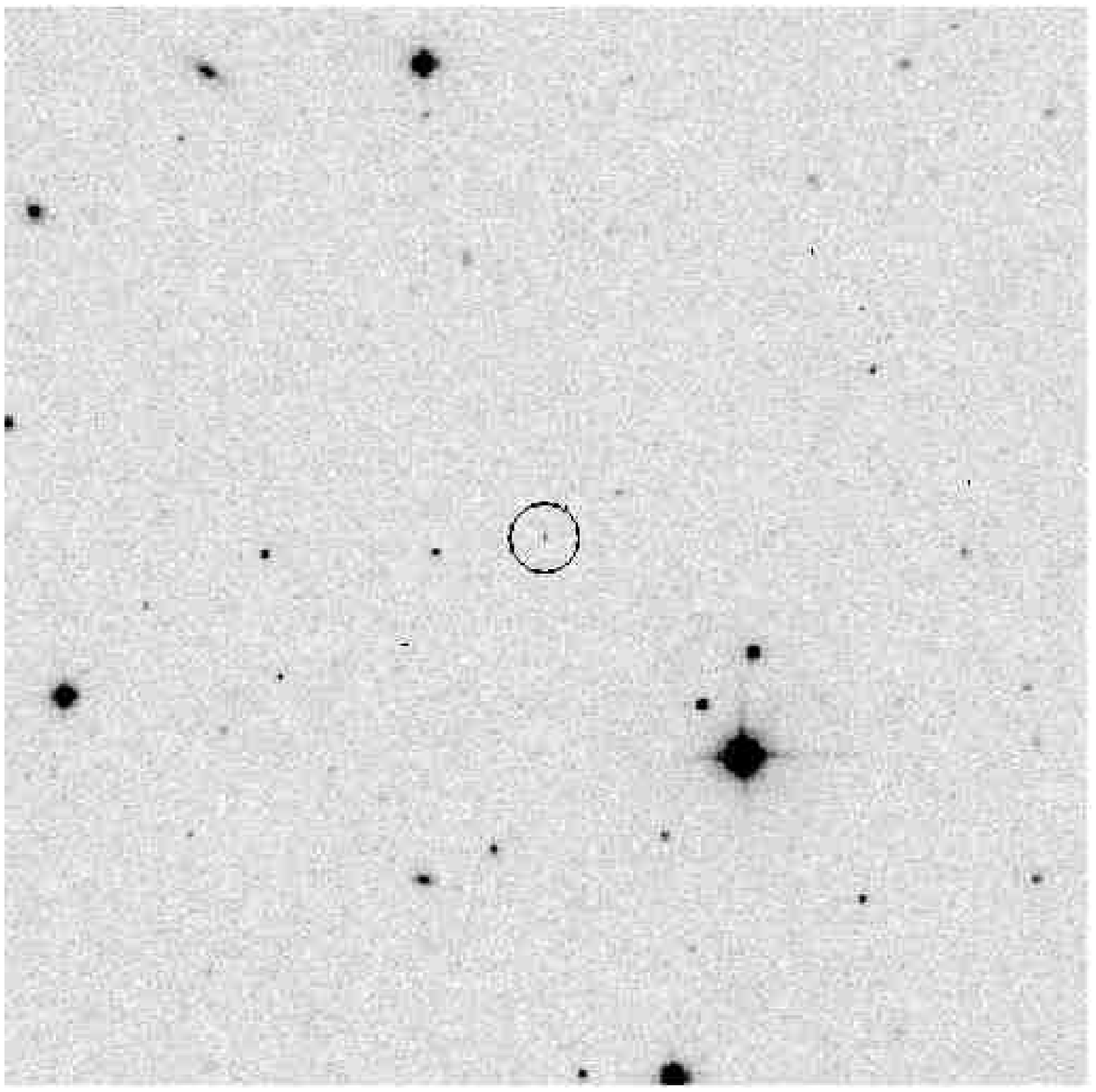}}
  \caption{$I$-band finder chart for 2MJ0013.}
  \label{t0657}
   \end{figure}

\begin{figure}
  \centering
  \resizebox{\hsize}{!}{\includegraphics[]{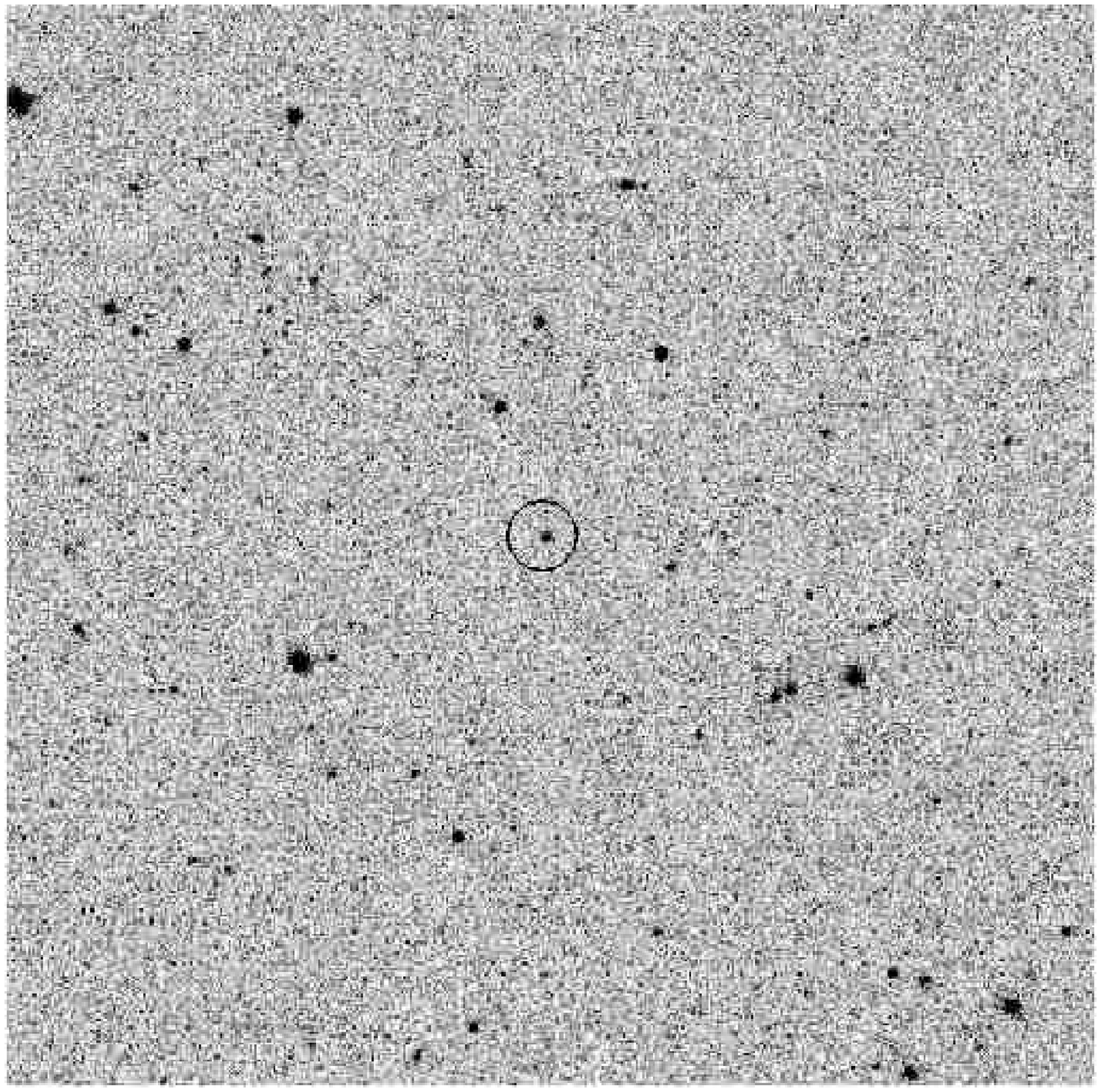}}
  \caption{$I$-band finder chart for 2MJ0032.}
  \label{d0779}
   \end{figure}

 \begin{figure}
  \centering
  \resizebox{\hsize}{!}{\includegraphics[]{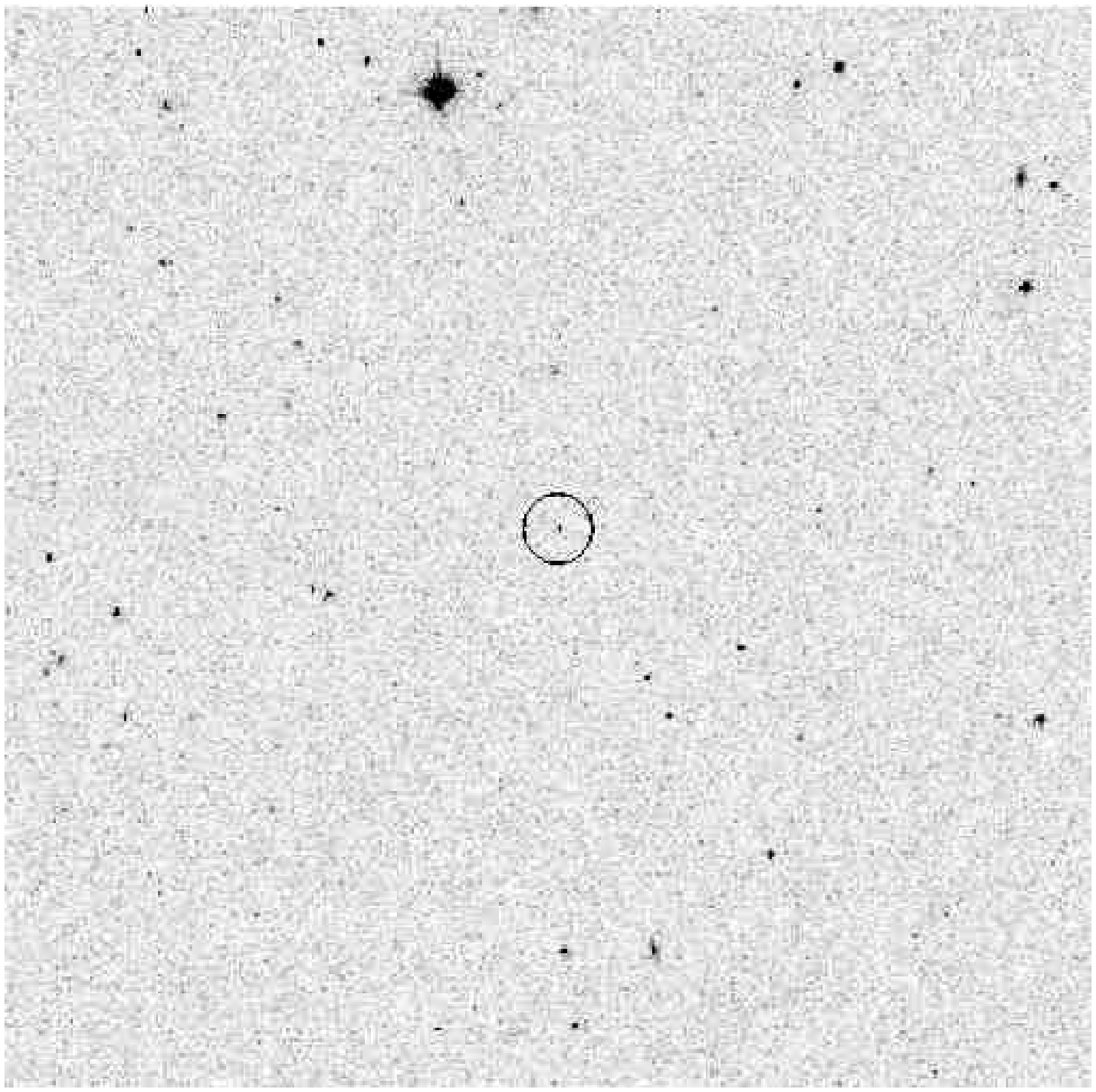}}
  \caption{$I$-band finder chart for 2MJ0034.}
  \label{d0785}
   \end{figure}

\begin{figure}
  \centering
  \resizebox{\hsize}{!}{\includegraphics[]{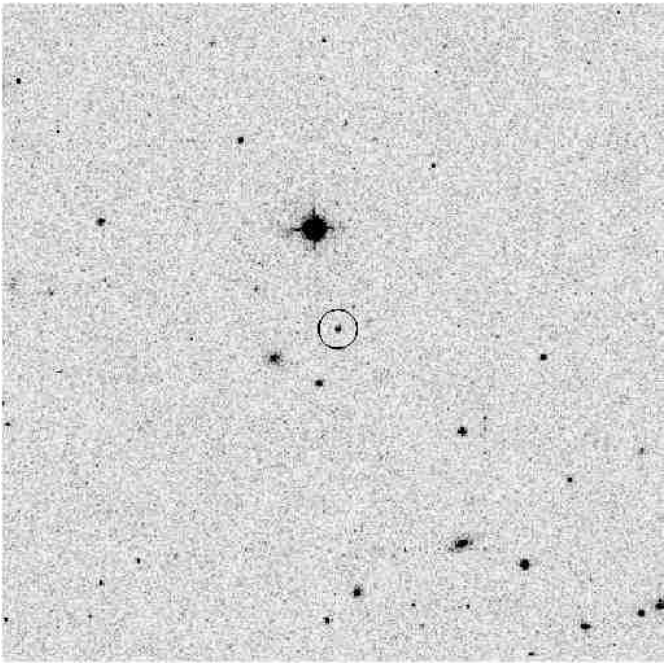}}
  \caption{$I$-band finder chart for 2MJ0428.}
  \label{d0799}
   \end{figure}

 \begin{figure*}
  \centering
  \resizebox{\hsize}{21cm}{\includegraphics[bb=0 70 543 790]{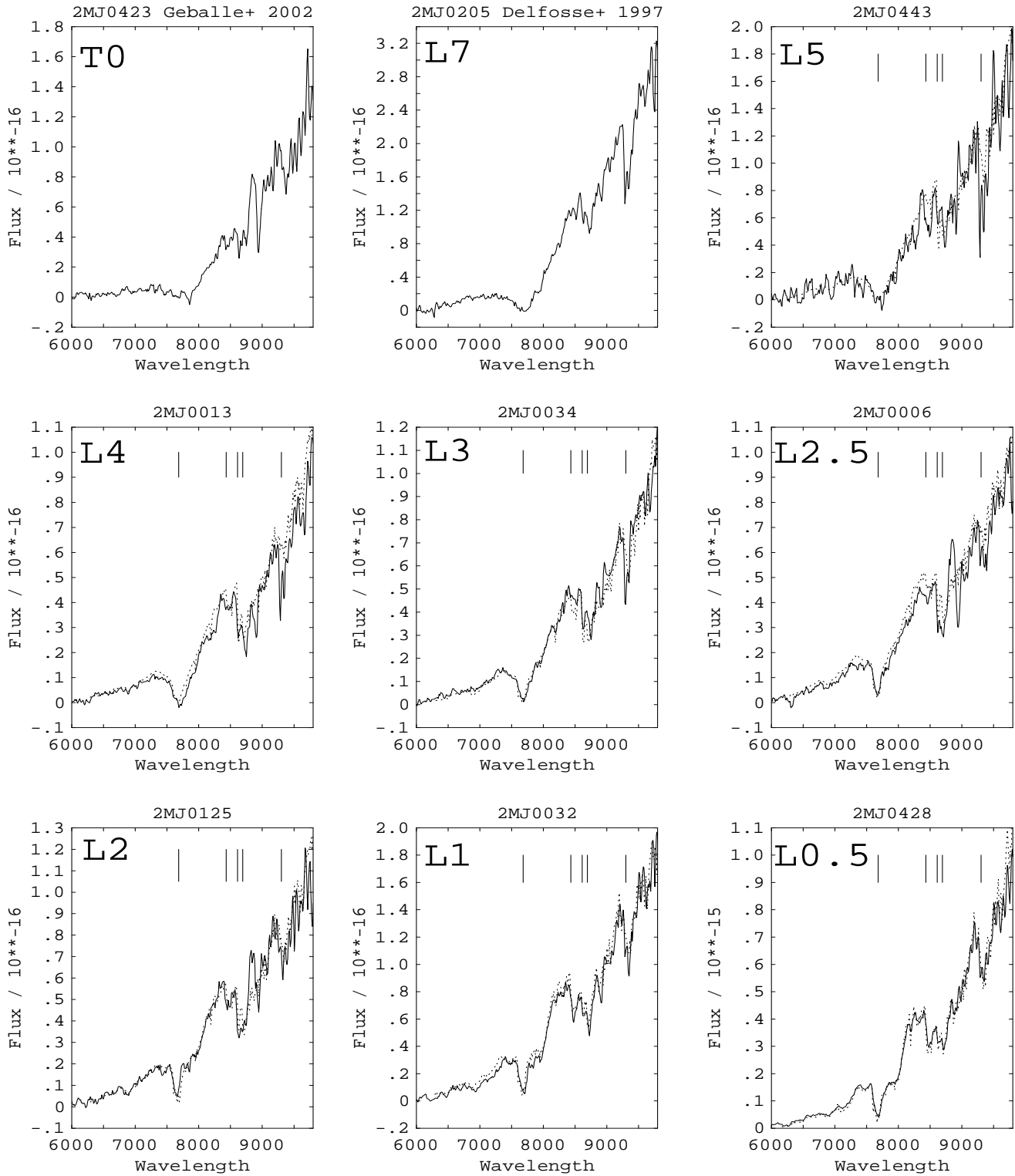}}
  \caption{Low resolution (30\AA) spectra of the 7 new objects, plus the 2 known objects SDSS J042348.57-041403.5 (2MJ0423; T0)
and DENIS-P0205.4-1159 (2MJ0205; L7). Spectra are overlain with L-type standards (dotted lines) of the given type from Kirkpatrick et al. (1999). Note 5 features visible at this resolution are marked by vertical bars in order of increasing wavelength. Note  K\,{\sc i} broadening to later types, the TiO band (head at 8423\,\AA) is visible in all early--L spectra and CrH and FeH bandheads at 8611 and 8692\,\AA~ respectively. The strong band near $\sim$\,9300\,\AA~ is telluric H$_2$O. Fringe residuals affect the spectra beyond 8500\,\AA~ at the 10--20\% level. Flux units are erg\,s$^{-1}$\,cm$^{-2}$\AA$^{-1}$.}
  \label{spt}
   \end{figure*}

\section{Results and Discussion}
 
Table 1 gives details of all 10 objects. We include 2MASS photometry for all the objects, and $I$ magnitudes
found by searching the SuperCosmos Sky Surveys database (\cite{ham01a};~\cite{ham01b}; \cite{ham01c}), which consists of digitised data derived from ESO and UKST
Schmidt plates. For the new objects, we draw attention to the high galactic latitude (col. 7 in Table 1) and include
spectral types and distances, for which details of the derivation will be given below. 

In Fig.~\ref{jhk}, we have plotted our seven new objects (filled squares) and 161 objects with errors
in $J-H$ and $H-K_s$ $\leq\pm$ 0.1 mag
from K02 (crosses) in the near-infrared colour-colour diagram. It is clear that the new objects
occupy the same region in this diagram as known L-dwarfs. Note that the new object 2MJ0428 is one of 
the $\sim$30 L-dwarfs brighter than $K_s\sim$\,12.2. In Fig.~\ref{ijhk} we compare the new L-dwarfs (filled squares) to a sample
from previous literature (for details see the caption) in the $I-J$, $H-K_s$ diagram. Note that here we exclude
DENIS-P J0205.4-1159 from the sample of \cite{del99} (open triangles) and instead plot our data for this object,
together with the other 2 previously known objects of Table 1 (filled triangles). $I-J$ has been shown to be
a good diagnostic of spectral type, increasing monotonically from late M to late T, although with more scatter
than the $I-z^{*}$ SDSS index (\cite{dah02}). Our
new objects occupy the region between $\sim$\,3\,$\leq$\,$I-J$\,$\leq$$\sim$\,3.7 which would be expected
for objects with spectral types between approximately L0 and L5, although one new object 2MJ0443 (L5)
seems anomalously blue for this spectral type. It is possible that the SuperCosmos $I$ photometry has been contaminated
by the presence of a faint galaxy (see Fig.~\ref{d0809}).

\subsection{Spectral type determinations}

The low resolution of our spectra (30\,\AA), intended only to allow differentiation between M and C-type
AGB stars, precludes the observation of narrow atomic lines in our L-dwarf spectra. However, as can be seen in Fig.~\ref{spt}, the spectra do exhibit broad K\,{\sc i} and features of CrH and FeH which immediately show
an L spectral type. The low resolution, and relatively poor signal-to-noise ratio of at least one of
the spectra (2MJ0443), preclude the use of spectral typing indices based on ratios of molecular and atomic features.
Instead, we have compared our spectra directly with those of L-type standards in the \cite{kir99} sample.
In Fig.~\ref{spt}, the spectra of the seven new L-dwarfs are overplotted with a representative standard
star (dotted lines) of the spectral-type indicated in the upper left corner of each panel. This spectral type is the one adopted for the new objects. The spectra of the known T0 and L7 objects SDSS J042348.57-041403.5 
and DENIS-P\,0205.4-1159 are also included in this figure for comparison.

When deriving the spectral types, each target spectrum was compared to at least one, and in many cases 
3 or 4 standards, including the representative standard we have chosen to plot. The main typing criteria at this
resolution were the width of the K\,{\sc i} resonance line, the strength of the TiO band at 8432\AA, and the
relative strengths of the CrH and FeH bandheads at 8611 and 8692\,\AA. These features are marked in Fig.~\ref{spt} by vertical bars (together with the broad band near 9300\,\AA~ which is due to H$_2$O, but which includes a significant telluric component). An additional diagnostic of spectral type is the slope of the spectrum near 8000\,\AA, which changes rapidly through very early L-types owing to the disappearance of VO. The spectral typing criteria used here are identical to the major ones used by \cite{kir99}. The excellent agreement
between the shape of our new spectra and the chosen standards in Fig.~\ref{spt} suggests the spectral types for the new objects are accurate to $\pm$\,1 subclass. 

Additionally, we have performed experiments using the pseudocontinuum
index PC3 of \cite{mar99}, which increases between early-M and late-L dwarf types but only
rather slowly in the early-L regime. The PC3 index compares the slope of the pseudocontinuum over 40\,\AA~ ranges centred at 8240\,\AA~ and 7560\,\AA, and the relations between it and spectral type for the ranges M2.5 to L1 and L1 to L6 are given by \cite{mar99}.  For the known T0 dwarf SDSS J042348.57-041403.5 (2MJ0423) these calibrations are not appropriate. However, for the other two late L-dwarfs with previously published types (2MJ0144
and 2MJ0205 in Table 1), we find agreement to within 1 subclass from our spectra using the PC3 index.

\begin{figure}
  \centering
  \resizebox{\hsize}{4in}{\includegraphics[]{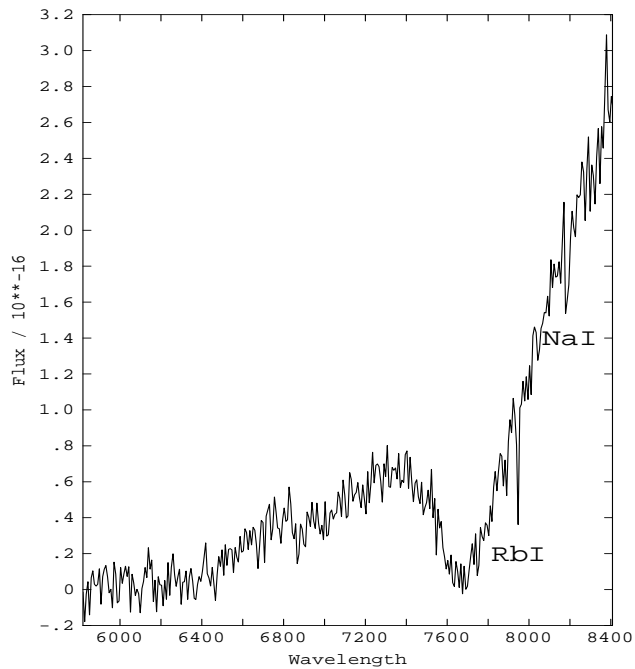}}
  \caption {2\AA~ resolution spectrum of the known object 2MASS J014435.3-071614 (2MJ0144) - spectral type L5 - rebinned to
8\AA~ resolution to clearly show the Rb\,{\sc i} line at 7948\AA~ and the Na\,{\sc i} doublet (unresolved at this resolution) at $\sim$ 8190\AA. The wavelength range does not include the hydride bands. Note the broad K\,{\sc i}. Flux units are erg~s$^{-1}$\,cm$^{-2}$\,\AA$^{-1}$.}
\label{lieb}
\end{figure}

For one object, 2MASS J014435.3-071614 (2MJ0144, classified L5 (Liebert et al. 2003), a higher resolution (2\,\AA) spectrum was obtained (see Fig.~\ref{lieb}), which does not include the hydride bands of FeH and CrH or the 8432\,\AA~TiO feature. After rebinning to 8\,\AA~ resolution to obtain a sufficient signal-to-noise ratio, we obtain a type of L4 (PC3\,=\,6.54). 

For the earlier L-type objects, to mitigate against of the very low resolution and moderate signal-to-noise ratio of the spectra, and the fact that the PC3 index is very slowly varying in this range, we increased the width of the 2 PC3 ranges to 100\,\AA. Spectral types of L1
were found for 2MJ0125, 2MJ0006, 2MJ0428 and 2MJ0032 (PC3\,=2.81, 2.69, 2.58 and 2.74 respectively). For 2MJ0034 we found L2 (PC3\,=\,3.34), and L3 for the remaining new objects 2MJ0013 and 2MJ0443 (PC3\,=\,4.39 and 4.74). In most cases the spectral types obtained using PC3 are systematically $\sim$ 1 subclass bluer than those found by direct visual comparison with standards. The exception is the lowest
signal-to-noise ratio spectrum 2MJ0443, where PC3 suggests L3 instead of L5. These discrepancies suggest that in order to use the PC3 index for these data, it would need to be recalibrated by observing spectral standards with the same instrumental setup. We note that, while the PC3 index is not satisfactory for classification here,
it indicates the consistency of the relative spectrophotometric calibration of the data. We adopt the spectral types
given by direct comparison with standards for the remainder of this discussion.

\begin{table*}
\begin{center}
\caption[]{Proper motion determinations for the 7 new objects. For 2 objects, the difference in
epoch between SuperCosmos and 2MASS positional information is too small for the derived motion to be significant. One measurement is clearly anomalously large, and the SuperCosmos position is unreliable because of blending. The penultimate column gives tangential velocity (km\,s$^{-1}$) computed using the distances of Table 1 and the proper motions added in quadrature.}
\begin{tabular}{llllll}\hline\hline
Name & $\mu_\alpha$cos$\delta$ & $\mu_\delta$ & $\Delta_{\rm epoch}$ & v$_{\rm trans}$ & Note \\
     &          mas\,yr$^{-1}$    &    mas\,yr$^{-1}$ & yr & km\,s$^{-1}$ & \\
\hline
2MJ0006 & 35 & 270 & 1.23 & 59 & $\Delta_{\rm epoch}$ small \\
2MJ0013 & 100 & 55 & 16.86 & 20 & - \\
2MJ0032 & 90 & 40 & 16.89 & 23 & - \\
2MJ0034 & 200 & 110 & 5.02 & 42 & - \\
2MJ0125 & 180 & 120 & 7.34 & 45 & - \\
2MJ0428 & $\sim$0 & 170 & 2.88 & 18 & $\Delta_{\rm epoch}$ small \\
2MJ0443 & 655 & 940 & 2.03 & 143 & anomalous\\
\hline
\end{tabular}
\end{center}
\end{table*}

\subsection{Distance derivations and proper motions}

We have used the M$_{J}$ - spectral type relations of \cite{haw02} to derive distances, given in Table 1. Two objects, 2MJ0428 (L0.5) and 2MJ0443 (L5) are rather
nearby ($\sim$ 30\,pc or less), as close as the 3 L-dwarfs discovered in the proper motion study of \cite{lod02}. We note here that the \cite{haw02} calibrations suggest the L7 object DENIS-P\,0205.4-1159 (2MJ0205) is very close (11\,pc). The T0 object SDSS J042348.57-041403.5 (2MJ0423) would also be nearby (12 pc), but we caution that the M$_J$ - spectral type relation is non-monotonic in the late-L to mid-T regime, as 
shown by Dahn et al. (2002), so may not be applicable to this object. 
Importantly, for the case of DENIS-P\,0205.4-1159, the distance is known {\it via} a trigonometric parallax determination to be 19.8\,pc, and the discrepancy between this and the 11\,pc distance derived from the spectral type is a result of this object
being an equal magnitude binary (\cite{dah02}). Given that our present sample is magnitude-limited, it is biased
towards finding such binaries: one object especially, 2MJ0428, which our results suggest to be rather close ($\sim$\,22\,pc) may also be a good candidate to investigate for duplicity. 

In Table 2, we have computed proper motions using the positional differences in $\alpha$ and $\delta$
between SuperCosmos and 2MASS images taken at different epochs. Col. 4 gives the timeline over which the proper motions were derived, and for three objects, we consider that it is too small for the proper motion to be 
reliable, and for one of these cases (2MJ0443) an anomalously large proper motion is derived. On examination of the
SuperCosmos UKST $I$ image for this case, we find that the image, which is resolved into the L-dwarf and a faint galaxy in the Danish $I$-band image of Fig.~\ref{d0809}, is clearly non-stellar and very indistinct, and it is possible that the SuperCosmos position
we have used is that of the nearby faint galaxy.

Moreover, we have derived transverse velocities using the spectroscopic distances of Table 1 and the
proper motion determinations. For the 2 cases in Table 2 for which $\Delta_{\rm epoch}$ is perhaps
unreliably small, we derive $\sim$\,60 and $\sim$\,20 km\,s$^{-1}$. However for the remaining four objects, we find $\sim$\, 20, 25,
40 and 45 km\,s$^{-1}$, which are quite consistent with the kinematics of disk stars. 

\section{Conclusions}

We have discovered seven new southern L-dwarfs at high galactic latitude serendipitously in a systematic survey for
distant halo AGB stars. We have derived spectral types in the range L0.5 -- L5 for the new objects by direct comparison to L-type spectral standards listed by \cite{kir99}. The seven new L-dwarfs have spectroscopic distances between $\sim$\,20 and 50\,pc.
Together with the proper motions we have derived by comparing 2MASS and SuperCosmos positions at different
epochs, we have determined transverse velocities consistent with the kinematics of disk stars. Follow-up
observations at higher spectral resolution are required to assess spectral types, and hence distances,
more accurately by using a larger number of spectral features and indices, and also to address the question of the new objects' masses, and potential substellarity, {\it via} observations of the Li\,{\sc i} 6708\,\AA~ line (\cite{reb96}).

\begin{acknowledgements}
TRK acknowledges financial assistance from the European Union Research Training Network `The Formation and Evolution of Young Stellar Clusters' (RTN1-1999-00436). NM thanks the ESO staff, in particular John Pritchard, for help in making the observations. KG thanks the Jumelage 18 "Astrophysique France-Arm\'{e}nie" for support. TRK thanks the referee, J. Davy Kirkpatrick, and Eduardo Mart\'{\i}n, for their comments and assistance. Part of this work was carried out during a visit of TRK to the Institute for Astronomy, University of Hawaii, funded by the National Science Foundation (NSF) grant AST-0205862. This publication makes use of data products from the Two Micron All Sky Survey, which is a joint project of the University of Massachusetts and the Infrared Processing and Analysis Center/California Institute of Technology, funded by the National Aeronautics and Space Administration and the national Science Foundation. This research has made use of the SIMBAD database, operated at CDS, Strasbourg, France.

\end{acknowledgements}

\end{document}